# Eco-Routing based on a Data Driven Fuel Consumption Model

**Xianan Huang & Huei Peng**
**University of Michigan, Ann Arbor, the United States**
E-mail : xnhuang@umich.edu

A nonparametric fuel consumption model is developed and used for eco-routing algorithm development in this paper. Six months of driving information from the city of Ann Arbor is collected from 2,000 vehicles. The road grade information from more than 1,100 km of road network is modeled and the software Autonomie is used to calculate fuel consumption for all trips on the road network. Four different routing strategies including shortest distance, shortest time, eco-routing, and travel-time-constrained eco-routing are compared. The results show that eco-routing can reduce fuel consumption, but may increase travel time. A travel-time-constrained eco-routing algorithm is developed to keep most the fuel saving benefit while incurring very little increase in travel time.

Topics/Intelligent Traffic Management and Control, Eco-Routing

## 1 INTRODUCTION

Ground transportation consumes 26.5% of the world energy in 2016 [1]. In 2014, 3.1 billion gallons of wasted fuel and 6.9 billion hours of extra time are caused by congestion [2]. Vehicle trip planning and routing based on traffic information and predicted fuel consumption can save fuel and travel time, the potential of which has not been deeply explored.

Intelligent transportation techniques have demonstrated promising results in reducing fuel consumption [3, 4]. An early study of eco-routing using average-speed-based fuel consumption model was conducted, showing 25% fuel saving compared with a shortest-time routing strategy [5]. To understand network-wide benefit, the user equilibrium and system optimal behavior were analyzed [6] and the authors concluded that the potential of fuel saving is 9.3%. Other factors such as signalized intersections [7] and penetration rate [8] were also studied.

A core piece for eco-routing algorithm development is a robust fuel consumption model. Microscopic fuel consumption models have been studied extensively [9], but for eco-routing, fuel consumption for large number of road sections is needed, thus fast computation is also needed. Macroscopic models [10] have also been studied to estimate fuel consumption without considering heterogeneity in driving, resulting in same fuel consumption for same average speed, thus not appropriate for eco routing. Mesoscopic models that use road link average speed and grade are widely used for eco-routing. By considering link-based variables, they can address driving heterogeneity, thus are more accurate than macroscopic models. However, most of the existing mesoscopic models for eco-routing are achieved with parametric regression-based models or power balance models, [11] and many are not accurate enough due to complexity of traffic scenario and nonlinearity of vehicle powertrains. Parametric methods like support vector machines (SVM) and neural networks (NN) were also studied [12]. Recently, a nonparametric model called multivariate adaptive regression spline (MARS) was studied [13]. MARS partitions the feature space into hypercubes with boundaries perpendicular to the axes of the feature space, thus a rotation of the coordinate axes can completely change the structure of the MARS model.

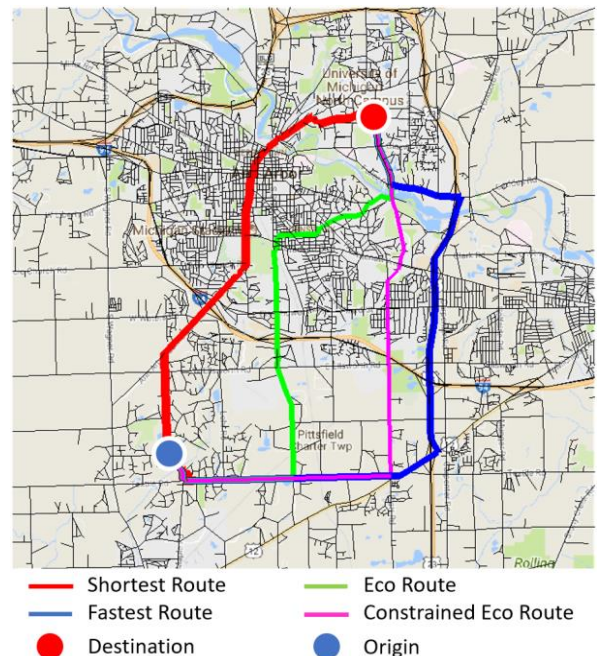

Fig. 1 Sample path for shortest route, fastest route, eco route, and constrained eco route with routing cost estimated from posted speed limit

The main idea of our method is that the fuel consumption model should (i) use credible physics-driven simulation model (like Autonomie [9] that we choose), (ii) the driving speed should be from real vehicle data which reflects the real-world operating condition of the road links, and (iii) instead of fitting individual trips, the model should aim to fit the expected value from many trips. We use the Gaussian Mixture Regression (GMR) to build our model [14]. The GMR models the joint density of model input and output, then derives the conditional expectation of the output from joint density function of the inputs and output, thus the model is invariant under any coordinate system. After the fuel consumption model is developed, we use it to evaluate the expected fuel consumption of different routing strategies including shortest-distance, shortest-time, eco-routing, and travel-time-constrained eco-routing. The main contributions of this paper include: 1) a nonparametric data driven fuel consumption model based on real-world driving data and Autonomie fuel consumption simulation; 2) a constrained eco-routing strategy addressing tradeoff between travel time and fuel consumption; and 3) studying the fuel consumption and travel time trade-off of different routing strategies.

The rest of the paper is organized as follows. The data used, the Autonomie model used, the Gaussian Mixture Regression model (GMR) and the constrained eco-routing method are presented in Section 2. Section 3 presents results and discussion. Conclusions and future work are given in Section 4.

## 2 METHODOLOGY
### 2.1 Data Description

The real-world travel speed and grade trajectories are obtained from the Safety Pilot Model Deployment (SPMD) database [15]. The SPMD program aims to demonstrate connected vehicle technologies. It has been recording naturalistic driving of up to 2,842 equipped vehicles, which is about 2% of total population in Ann Arbor, Michigan for more than three years. As of April 2016, 56.2 million kilometers have been logged, making SPMD one of the largest naturalistic driving databases. The query criteria used for this paper are as follows:

− From May 2013 to October 2013
− All passenger car
− Trip duration longer than 10 minutes
− Trip distance longer than 300 meters
− Trips in the Ann Arbor area: latitude between 42.18º and 42.34º, and longitude between -83.85º and -83.55º

The queried results include 321,945 trips, which cover a total distance of more than 3.7 million kilometers and total time of more than 93,926 hours from 2,468 drivers. The data covers 9,745 of the 11,506 links in the Ann Arbor area, with 5,599 links covered by more than 100 trips. The links with more than 100 trips are shown in Fig. 2, which consist of major roads, minor roads, ramps, and highway sections.

Table 1 Key vehicle parameters for Autonomie microscopic simulation

| Vehicle Mass [kg] | 1,246 |
| Max Engine Power [kW] | 178.7 |
| Max Engine Efficiency [%] | 36 |
| Max Engine Speed [rad/s] | 628.2 |
| Idle Engine Speed [rad/s] | 62.8 |
| Transmission Gear Number | 6 |
| Fuel Type | Gasoline |

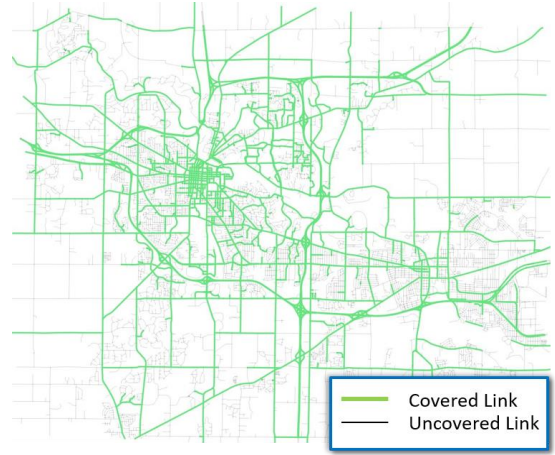

Fig. 2 Links with more than 100 trips each from the queried data

The speed and grade trajectories are used as the inputs to Autonomie [9], a microscopic fuel consumption model developed by the Argonne National Lab. The key vehicle parameters simulated are listed in Table 1. We assume the target vehicle is a mid-sized gasoline engine vehicle. Including multiple vehicles and powertrain types will be considered in the future work.

### 2.2 Fuel Consumption Model

We use Autonomie output as the ground truth to develop our fuel consumption model, which fits the average fuel consumption of all trips on all road links in Ann Arbor. In the modeling framework, we treat speed limit as a categorical variable and fit a distinct set of model parameters for links with different speed limits. The fuel consumption model for each speed limit category is obtained using the Gaussian Mixture Regression model (GMR) technique. Instead of modeling the regression function directly, GMR models the joint distribution of input and output variables and get the regression function through conditional distribution of the output as functions of the inputs. We denote the input variable as $X = [x_1, \ldots, x_i, \ldots, x_P] \in R^{N \times P}$, where $x_i \in R^N$ is individual input variable, $N$ is the sample size, $P$ is the number of input variables, and $Y$ is the output variable, i.e., fuel consumption. The optimal model parameters are obtained by solving

$$\theta^* = argmin_\theta \|Y - f(x, \theta)\| \quad (1)$$

where $f(x, \theta)$ is the modeled regression function. The objective of the optimization problem is to minimize the norm of the regression error, which is equivalent to maximize the conditional likelihood of the output on the input variables

$$\theta^* = argmax_\theta \prod_{i=1}^{N} p(Y_i | X_i, \theta) \quad (2)$$

The joint distribution of input and output can be factorized as

$$P(Y, X | \theta) = P(Y | X, \theta) P(X | \theta) \quad (3)$$

Since $P(X|\theta)$ depends only on the input variable and thus is independent of $\theta$, maximize conditional likelihood of output is equivalent to maximize joint likelihood of input and output.

$$\theta^* = argmax_\theta \prod_{i=1}^{N} P(Y_i, X_i | \theta) \quad (4)$$

In GMR, the joint distribution is modeled as gaussian mixture model (GMM).

$$f_{X,Y}(x,y) = \sum_{k=1}^{K} \pi_k f_{X,Y,k}(x,y) \quad (5)$$

$$f_{X,Y}(x,y) = \sum_{k=1}^{K} \pi_k f_{Y|X,k}(y|x) f_{X,k}(x) \quad (6)$$

where $f_{X,Y}(x,y)$ is the overall joint density function, $\pi_k$ is the mixing coefficient for each component, $f_{X,Y,k}(x,y)$ is the joint density for each component, which follows a multivariate Gaussian distribution. For each component of GMM, the conditional distribution of output on the input still follows Gaussian distribution and can be presented in a closed form. The marginal distribution of $X$ is

$$f_X(x) = \int f_{X,Y}(x,y) dy = \sum_{k=1}^{K} \pi_k f_{X,k}(x) \quad (7)$$

Thus, the conditional density of output is

$$f_{Y|X}(y|x) = \sum_{k=1}^{K} w_k(x) f_{Y|X,k}(y|x) \quad (8)$$

where the posterior of component probability $w_k(x)$ is obtained from marginal distribution of $X$.

$$w_k(x) = \frac{\pi_k f_{X,k}(x)}{\sum_{k=1}^{K} \pi_k f_{X,k}(x)} \quad (9)$$

To obtain parameters of the GMM for the joint density, the most popular approach is to apply the Expectation-Maximization (EM) algorithm and use the maximum likelihood method. In the Expectation (E) step, the mixing coefficient is estimated using the mean and covariance of each component by calculating the posterior; in the Maximization (M) step, the mean and covariance are estimated from the maximum likelihood method using the mixing coefficient from the E step. To apply the EM algorithm, one needs to specify the component number of the GMM, which can be achieved through cross validation. However, since we have multiple sets of parameters due to the categorical variable (road link speed limit), specifying component number for each speed limit through cross validation can be time consuming. Thus, instead of the EM algorithm, we adopt the Bayesian modeling framework, which models the parameters as hidden random variables and inference the expectation of the parameters from the data [16]. Multiple approaches can be used to solve the inference problem, including Markov Chain Monte Carlo (MCMC) and Variational Inference (VI). We apply the VI approach to get the expected values of the parameters. The approach is summarized as follows. Denote $\tilde{X} = [X, Y]$ as joint of input and output, $Z = \{z_{nk}\}_{N \times K}$ as the indicator variable of the component for each data point, which is a binary variable.

$$P(Z|\pi) = \prod_{n=1}^{N} \prod_{k=1}^{K} \pi_k^{z_{nk}} \quad (10)$$

The parameters are modeled with their corresponding conjugate priors, i.e., Dirichlet distribution for $\pi$ and Gaussian-Wishart distribution for mean and covariance.

$$P(\pi) = Dir(\pi|\alpha_0) = C(\alpha_0) \prod_{k=1}^{K} \pi_k^{\alpha_0 - 1} \quad (11)$$

$$P(\mu, \Sigma) = P(\mu|\Sigma) P(\Sigma)$$

$$= \prod_{k=1}^{K} N(\mu_k|m_0, \beta_0 \Sigma_k) W(\Sigma_k^{-1}|W_0, v_0) \quad (12)$$

where $\alpha_0, m_0, \beta_0, W_0, v_0$ are hyperparameters. The hidden variables to inference includes the indicator variable $Z$ and the model parameters $\pi, \mu, \Sigma$. The joint distribution is factorized as

$$P(\tilde{X}, Z, \pi, \mu, \Sigma) = P(\tilde{X}|Z, \pi, \mu, \Sigma) P(Z|\pi) P(\pi, \mu, \Sigma) \quad (13)$$

The VI approach uses a tractable (factorizable in this case) posterior distribution of the hidden variables to approximate the original posterior distribution and minimize the Kullback-Leibler (KL) divergence between the true distribution and the approximated distribution. The approximate distribution is

$$q(Z, \pi, \mu, \Sigma) = q(Z) q(\pi, \mu, \Sigma) \quad (14)$$

It can be shown that the stationary point of the KL divergence minimization problem satisfies

$$\ln q^*(Z) = E_{\pi, \mu, \Sigma}(\ln p(\tilde{X}, Z, \pi, \mu, \Sigma)) + const \quad (15)$$

$$\ln q^*(\pi, \mu, \Sigma) = E_Z(\ln p(\tilde{X}, Z, \pi, \mu, \Sigma)) + const \quad (16)$$

From the stationary point condition, we can update $q(Z)$ and $q(\pi, \mu, \Sigma)$ alternatively and iterate until convergence. The algorithm is initialized with hyperparameters of prior distributions. The approximated posterior of $Z$ is first updated through (15), the mean and covariance are then obtained using the maximum likelihood method. For more details, one can refer to chapter 10 of [16].

The expectation of the mixing coefficient is

$$E(\pi_k) = \frac{\alpha_0 + N_k}{K\alpha_0 + N} \quad (17)$$

For a component with small sample size, $N_k \approx 0$, if a small hyperparameter $\alpha_0$ is used, as sample size approaches infinity

$$\lim_{N \to \infty} E(\pi_k) = \lim_{N \to \infty} \frac{\alpha_0 + N_k}{K\alpha_0 + N} = 0 \quad (18)$$

Thus, a small hyperparameter for mixing coefficient can be used to remove the redundant components. In this way, we don't need to specify the component number for GMM. As the sample size increases, the influence of hyperparameters decreases. To see this, take mixing coefficient for example, since $\alpha_0$ and $K$ are finite, as $N$ and $N_k$ approaches infinity, the expectation is determined by the total sample size and the sample size for each component. Thus, the algorithm is less sensitive to tuned parameters compared with other algorithms such as SVM and neural network.

Table 2 Input variables for fuel consumption model

| | |
|---|---|
| Motion Related | Average Speed [m/s] |
| | Speed Change [m/s] |
| Link Related | Average Grade [rad] |
| | Link Length [m] |
| | Posted Speed Limit [m/s] |

The input variables we use for the fuel consumption model are listed in Table 2. We include both linear and the 2nd order terms, including cross-coupling 2nd order terms of the input variables. Since we treat the speed limit as a categorical variable, with the assumption that free flow speed can be approximated by the speed limit, average speed is also an indicator of the congestion status. Speed change and average grade are included to capture the kinetic and potential energy change.

### 2.3 Constrained Eco-Routing

To evaluate the benefit of eco-routing, we developed a travel-time-constrained eco-routing strategy. In this study, we define the links as nodes in a routing graph, and two nodes are connected by a directed edge if there exists movement allowing traveling from one link to its adjacent link. By using this definition, we can include speed change as part of the action cost to evaluate the expected fuel consumption. In this problem, we model all links as directed and do not allow U-turn. The algorithm is based on dynamic programming [17], which solves the optimization problem recursively based on the Bellman optimality principle.

$$x_i^* = \mathrm{argmin}_{x_i \in adj_{out}(x_{i-1})} g(x_i, x_{i-1}) + f^*(x_{i-1}) \quad (19)$$

$$f^*(x_i) = \min_{x_i \in adj_{out}(x_{i-1})} g(x_i, x_{i-1}) + f^*(x_{i-1}) \quad (20)$$

$$f^*(x_d) = 0 \quad (21)$$

where $x_i$ is the optimal next link location, $x_{i-1}$ is the last link location. The next links should be in the adjacent set of last link. $f^*(x_{i-1})$ is the optimal value function of the last link. $g(x_i)$ is the transition cost defined as the weighted sum of travel time and fuel consumption in travel-time-constrained eco-routing. $f^*(x_d)$ is value function associated with the destination link, and is defined to be 0. The transition cost is defined as

$$g(x_i) = (1 - w_t)c(x_i, x_{i-1}) + w_t t(x_i) \quad (22)$$

where $c(x_i, x_{i-1})$ is the expected fuel consumption and $t(x_i)$ is the expected travel time for the current link. To address the travel time constraint, a soft constraint is defined with respect to time limit $t_c$. The soft constraint is achieved through weighting parameter between fuel consumption cost and travel time cost $w_t$. The soft constraint is modeled with a sigmoid function as shown in Fig. 3, where the travel time limit is calculated from

$$t_c(x_i) = (1 + \epsilon)t^*(x_i) \quad (23)$$

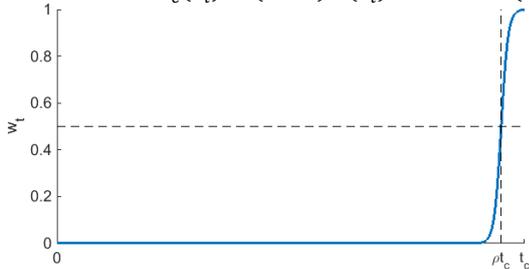

Fig. 3 Weighting parameter for travel time

where $\epsilon$ is a constant and $t^*(x_i)$ is the travel time of the shortest time solution from the destination to link $x_i$. The travel time constraint indicates that we only allow the travel time to increase no more than a certain percentage compared with the travel time of the fastest route. For shortest time routing and unconstrained eco-routing, we define $w_t$ in (22) to be 1 and 0 respectively. For shortest distance routing, we define the transition cost to be the link length.

### 2.4 Travel Demand Identification

To estimate the expected fuel consumption and travel time for different routing algorithms, we use travel origin-destination pairs from real-world driving data. We assume that the number of vehicles using proposed routing algorithm is limited, i.e., the vehicles cannot cause notable change to the travel speed of the links in the traffic network. The data to estimate travel demand is during May 2013 to October 2013, from 17:00 to 19:00 on weekdays. 25,001 trips were identified within the specified time. The origin and destination locations are identified through a density based cluster algorithm called OPTICS [18]. The advantage of this algorithm compared with other distance based clustering algorithms such as DBSCAN [19] is that it can cluster data with density change. This is critical in our analysis since the spatial densities of trip origin and destination locations can be affected by multiple factors such as parking lot size. We only include trips happening at least once per week. There are 3,031 frequently visited origin-destination pairs identified, and the identified starting and ending locations are shown in Fig. 4.

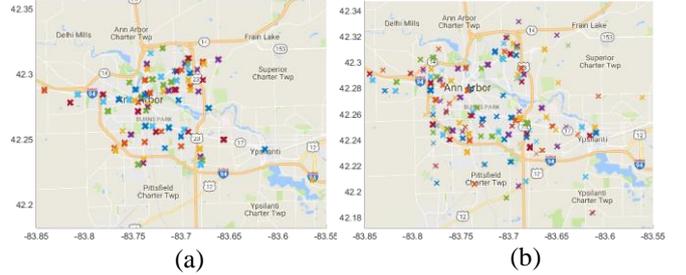

Fig. 4 Trip locations identified with OPTICS: (a) Trip starting locations; (b) Trip ending locations

## 3 RESULTS AND DISCUSSION

### 3.1 Fuel Consumption Model Performance

The fuel consumption model performance is measured using coefficient of determination ($R^2$) and mean absolute percent error (MAPE). Since the objective of the model is to predict the conditional expectation of fuel consumption on motion and link variables, we compare the model output with the conditional expectation of fuel consumption given the average speed and speed change. To get the conditional expectation, we fit individual GMR for each link with more than 100 trips. Through the model of individual link, we can get the conditional expectation of fuel consumption as the complete model described in Section 2.1. We randomly selected 70% links with more than 100 events as training dataset, and the rest as verification dataset. We use the conditional expected fuel consumption of test dataset as the ground truth. We compared our model with several benchmarks including the average speed model [5], the power balance model which is the foundation of MOVES [10], and the neural network model. Parameters of the benchmark models are also estimated from training dataset. For the neural network model, we used a two-layer structure with two fully connected layers, and sigmoid function as the activation function for the output of layer 1. The relative error histograms of the models are shown in Fig. 5 and model performance metrics are summarized in Table 3.

From the histogram and performance metrics, we can see that both our GMR model and the neural network model have superior performance over the other two models. Neural network models with well-tuned structure and parameters including number and structure of layers, type of activation functions, and number of hidden variables have the potential to achieve similar or better performance compared with our GMR model. However, the major advantage of our model is that it is nonparametric while many parameters need to be tuned for neural network models. Also, the final form of our model is simpler and thus should be more robust compared with neural network based models.

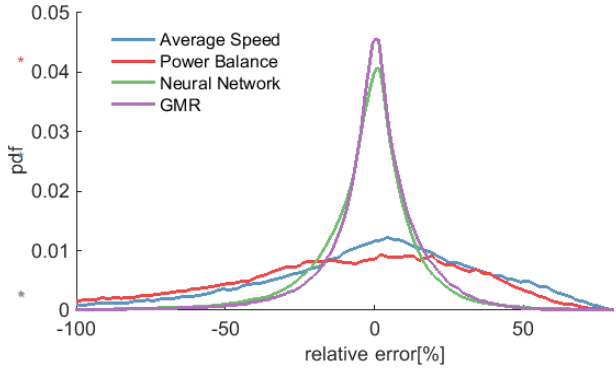

Fig. 5 Error histograms of fuel consumption models

Table 3 Performance of fuel consumption models

| Model | $R^2$ | MAPE [%] |
|---|---|---|
| Average speed model | 0.77 | 37.63 |
| Power balance model | 0.86 | 46.22 |
| Neural Network | 0.98 | 15.60 |
| **GMR** | **0.98** | **10.08** |

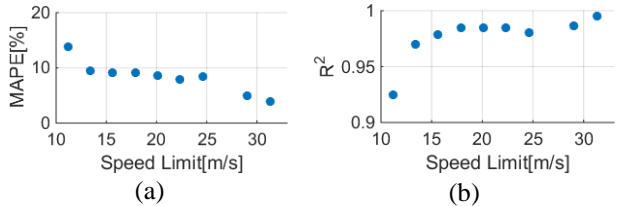

Fig. 6 Model performance for different speed limits: (a) MAPE; (b) $R^2$

The GMR model performance for links with different speed limit are shown in Fig. 6. The worst performance happens at links with low speed limit 11.18 m/s (25 mph) with MAPE 13.78%. The MAPE for links with higher speed limits are less than 10%. The reason, we believe, is that links with lower speed limit contain more speed and traffic variation. Also, at low speed and low torque, the engine fuel consumption is highly nonlinear with power, while for high power operation the fuel consumption – power relation is more linear.

### 3.2 Routing Results

The routing algorithm is applied to the 3,031 identified frequent OD pairs as described in Section 2.4. The studied Ann Arbor traffic network consists of 21,569 directed links with variate link types including local, minor, major, collector, ramp, and highway. The computation time to solve all-to-one routing result is around 13 s on a computer with Intel Core i7 and 16 G RAM. Considering requirement for the travel time of shortest-time routing, the computation time for constrained eco-routing is about 26 s. The routing cost are evaluated based on historical average speed during the studied hours. The uncovered links are imputed with their posted speed limits. Since they are never traveled by the sample vehicles over 6 months, we hypothesize these links are less traveled and the posted speed limit is a reasonable approximation for the free flow speed. To get the historical average speed, we use GMM to approximate average speed distribution of individual links and estimate the posterior of mixing coefficient based on speed during the sampled hours. The expectation of travel speed is estimated with the estimated posterior of the mixing coefficient.

To compare travel time and fuel consumption for different routing strategies, travel time and fuel consumption of different strategies are normalized with the travel time of fastest route and the fuel consumption of unconstrained eco-route respectively. The normalized costs are shown in Fig. 7. The scatter plot is overlaid with expectation of cost estimated with the OD pair travel frequency. The error bars for each routing solution are 10% and 90% percentiles respectively. The expectation of travel time and fuel consumption are summarized in Table 4.

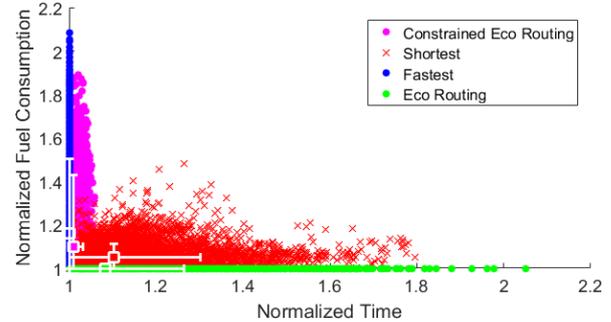

Fig. 7 Normalized travel time and fuel consumption for different routing strategies

Table 4 Expected travel time and fuel consumption of different routing strategies

|  | Fuel consumption [kg] | Travel Time [s] |
|---|---|---|
| Shortest | 0.4809 | 611.37 |
| Fastest | 0.5312 | 554.45 |
| Eco-routing | 0.4576 | 601.04 |
| Constrained eco-routing | 0.5038 | 559.49 |

From the results, we can see that the shortest path consumed less fuel compared with the fastest path algorithm, while the travel time is increased significantly. Also, with a maximum of 6.48% increase in travel time, the constrained eco-routing solution has expected fuel saving of 5.16% and maximum saving of 51.8%, compared with the fastest-path solution. It's also noted that for the given OD pairs, 28% of the eco-routing solution are the same as the fastest-path solution, and 27% is the same as the shortest-path solution. For constrained eco-routing results, 55% is the same as the fastest-route solution and 27% is the same as the shortest-path solution. Besides that, 28% of shortest path and fastest-path are the same. The difference between eco-routing and constrained eco-routing is due to the travel time constraints. To see the influence of traffic status on the routing results, we normalize the results of different strategies obtained with historical link travel speed with the travel time of fastest route and the fuel consumption of unconstrained eco-route under free traffic condition for which routing costs are estimated with the posted speed limits. The results are shown in Fig. 8.

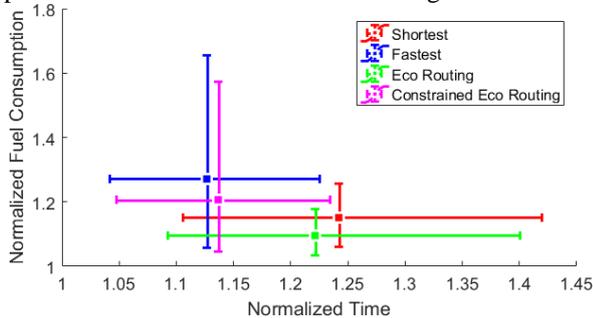

Fig. 8 Travel time and fuel consumption obtained with historical cost normalized with results from routing results of posted speed limit

The results show that with congestion formed during rush hour, travel time and fuel consumption are increased compared with free flow case. Shortest-path results in largest increment with 14.12% in time and 10.04% in fuel due to lack of consideration of average speed change. Fastest-path has 12.67% increment in travel time and 8.74% increment in fuel. With fuel consumption as part of the objective function, constrained eco-routing increased fuel consumption by 6.96%, which is the lowest one compared with others, but the travel time is increased by 13.01%, which is larger than fastest-path solution, but still smaller than shortest-path solution. Therefore, an accurate average speed estimation model using real-time information from intelligent transportation system such as connected vehicles and intelligent traffic light can play an significant role in eco-routing.

## 4 CONCLUSION AND FUTURE WORK

A nonparametric fuel consumption model is developed to estimate expected link fuel consumption conditional on motion and link variables. The model parameters are estimated from a large scale connected vehicle test database with simulated fuel consumption from the Autonomie software. The model is used to calculate constrained eco-routing results, which are found to save 5.16% fuel while incurring 0.91% travel time increase, compared with the fastest-path solution of frequent original-destination pairs of the Ann Arbor road network. Next step of this research includes to consider a wider set of vehicle parameters and different powertrains and to develop real-time constrained eco-routing implementations.

## 5 ACKNOWLEDGEMENT

The research is supported by U.S. Department of Energy under the award DE-EE0007212. We would like to thank Dr. Joshua Auld from the Argonne National Lab for providing Autonomie and the software for map-matching of GPS traces. We would also like to thank Dr. James R. Sayer and Ms. Debra Bezzina for accessing the Safety Pilot Model Deployment data.

## REFERENCES

[1] EIA, "Annual Energy Outlook 2017 with projections to 2050," pp. 1–64, 2017.
[2] D. Schrank., B. Eisele., T. Lomax., and J. Bak., "2015 Urban Mobility Scorecard," *Texas A&M Transp. Institue*, vol. 39, no. August, p. 5, 2015.
[3] X. Huang and H. Peng, "Speed Trajectory Planning at Signalized Intersections Using Sequential Convex Optimization," pp. 2992–2997, 2017.
[4] H. a. Rakha, K. Ahn, and K. Moran, "INTEGRATION Framework for Modeling Eco-routing Strategies: Logic and Preliminary Results," *Int. J. Transp. Sci. Technol.*, vol. 1, no. 3, pp. 259–274, 2012.
[5] K. Boriboonsomsin, M. Barth, S. Member, W. Zhu, and A. Vu, "ECO-Routing Navigation System based on Multi-Source Historical and Real-Time Traffic Information," *Network*, vol. 13, no. 4, pp. 1694–1704, 2012.
[6] K. Ahn and H. A. Rakha, "Network-wide impacts of eco-routing strategies: A large-scale case study," *Transp. Res. Part D Transp. Environ.*, vol. 25, pp. 119–130, 2013.
[7] J. Sun and H. X. Liu, "Stochastic Eco-routing in a Signalized Traffic Network," *Transp. Res. Procedia*, vol. 7, pp. 110–128, 2015.
[8] L. Guo, S. Huang, and A. W. Sadek, "An Evaluation of Environmental Benefits of Time-Dependent Green Routing in the Greater Buffalo–Niagara Region," *J. Intell. Transp. Syst.*, vol. 17, no. 1, pp. 18–30, 2013.
[9] A. Rousseau, P. Sharer, and F. Besnier, "Feasibility of Reusable Vehicle Modeling: Application to Hybrid Vehicles," *SAE Tech. Pap.*, no. 2004-01–1618, p. 12, 2004.
[10] J. Kwon, A. Rousseau, and P. Sharer, "Analyzing the uncertainty in the fuel economy prediction for the EPA MOVES binning methodology," *SAE Int.*, 2007.
[11] M. Kubicka *et al.*, "Performance of current eco-routing methods," *IEEE Intell. Veh. Symp. Proc.*, vol. 2016–Augus, pp. 472–477, 2016.
[12] W. Zeng, D. Candidate, T. Miwa, and T. Morikawa, "Application of machine learning and heuristic k- shortest path algorithm to eco-routing problem with travel time constraint," pp. 1–18, 2016.
[13] Y. Chen, L. Zhu, J. Gonder, S. Young, and K. Walkowicz, "Data-driven fuel consumption estimation: A multivariate adaptive regression spline approach," *Transp. Res. Part C Emerg. Technol.*, vol. 83, pp. 134–145, 2017.
[14] H. G. Sung, "Gaussian Mixture Regression and Classification," Rice University, 2004.
[15] D. Bezzina and J. Sayer, "Safety pilot model deployment: Test conductor team report," *Rep. No. DOT HS*, vol. 812, no. June, p. 171, 2014.
[16] C. M. Bishop, *Pattern Recognition and Machine Learning*. 2006.
[17] D. P. Bertsekas, "Dynamic Programming and Optimal Control 3rd Edition , Volume II by Chapter 6 Approximate Dynamic Programming Approximate Dynamic Programming," *Control*, vol. II, pp. 1–200, 2010.
[18] M. Ankerst, M. M. Breunig, H.-P. Kriegel, and J. Sander, "Optics: Ordering points to identify the clustering structure," *ACM Sigmod Rec.*, pp. 49–60, 1999.
[19] M. Ester, H. P. Kriegel, J. Sander, and X. Xu, "A Density-Based Algorithm for Discovering Clusters in Large Spatial Databases with Noise," *Proc. 2nd Int. Conf. Knowl. Discov. Data Min.*, pp. 226–231, 1996.